\documentclass[pmlr]{jmlr}% new name PMLR (Proceedings of Machine Learning Research)

\usepackage{indentfirst}
\usepackage{caption}

\usepackage{longtable}% for long tables

\usepackage{booktabs}
\usepackage[load-configurations=version-1]{siunitx} % newer version
\theorembodyfont{\upshape}
\theoremheaderfont{\scshape}
\theorempostheader{:}
\theoremsep{\newline}

\jmlrworkshop{CAWS 2023}

\title{Linkages among the Foreign Exchange, Stock, and Bond Markets in Japan and the United States}

  \author{\Name{Yi Jiang} \Email{s7022101@st.shiga-u.ac.jp}\\
  \addr Graduate School of Data Science, Shiga University, Japan\\
  \Name{Shohei Shimizu} \Email{shohei-shimizu@biwako.shiga-u.ac.jp}\\
  \addr Graduate School of Data Science, Shiga University, Japan\\
  Center for Advanced Intelligence Project, RIKEN, Japan}

\begin{document}

\maketitle

\begin{abstract}
While economic theory explains the linkages among the financial markets of different countries, empirical studies mainly verify the linkages through Granger causality, without considering latent variables or instantaneous effects. Their findings are inconsistent regarding the existence of causal linkages among financial markets, which might be attributed to differences in the focused markets, data periods, and methods applied. Our study adopts causal discovery methods including VAR-LiNGAM and LPCMCI with domain knowledge to explore the linkages among financial markets in Japan and the United States (US) for the post Covid-19 pandemic period under divergent monetary policy directions. The VAR-LiNGAM results reveal that the previous day’s US market influences the following day’s Japanese market for both stocks and bonds, and the bond markets of the previous day impact the following day’s foreign exchange (FX) market directly and the following day’s Japanese stock market indirectly. The LPCMCI results indicate the existence of potential latent confounders. Our results demonstrate that VAR-LiNGAM uniquely identifies the directed acyclic graph (DAG), and thus provides informative insight into the causal relationship when the assumptions are considered valid. Our study contributes to a better understanding of the linkages among financial markets in the analyzed data period by supporting the existence of linkages between Japan and the US for the same financial markets and among FX, stock, and bond markets, thus highlighting the importance of leveraging causal discovery methods in the financial domain.
\end{abstract}

\section{Introduction}
%\label{sec:intro}

The interdependence among the FX, stock, and bond markets arising from the intricate relationship between FX rates, interest rates, and stock prices in different countries is an extensively accepted phenomenon \citep{eicher2009international}. Most empirical studies explain market linkages via Granger causality, which may not provide precise and rigorous conclusions regarding the existence of unobserved variables or instantaneous effects. The recently developed causal discovery approach \citep{spirtes2000causation} \citep{glymour2019review} offers a better solution; however, it has rarely been utilized in this field. Moreover, the extant research \citep{saji2022stock} \citep{deaton2020jpy} is primarily based on data before 2019, which means that important changes in the economic environment and monetary policies post Covid-19 pandemic have not been considered, despite having influenced the market linkages. In this study, we leverage causal discovery models to elucidate market linkages among FX, stock, and bond markets in Japan and the US between July 2021 and December 2022.

Several economic and financial theories aim to shed light on market linkages. One such theory is the Interest Rate Parity (IRP) theory \citep{aliber1973interest}, which postulates that the difference in interest rates on bonds between two countries should be equivalent to the gap between the forward and spot FX rates. IRP emphasizes the linkage between FX and bond markets. The capital market theory \citep{fama1970efficient} asserts that the bond market is closely linked to stock markets. In an economy with a low interest rate, stocks tend to become more attractive as investment products, leading to an increase in the stock market. Although economists have theoretically explained the market linkages, the real world operates in a much more complex manner. Financial markets are affected by multiple factors and the linkages among them change based on global economic conditions, geopolitical events, and other factors.

This study examines the financial markets of Japan and the US because these two countries’ markets are regarded as closely connected through significant trade and investment flows. According to the data provided by the Ministry of Finance, Japan's exports to the US amounted to 18,255 billion Japanese Yen (JPY) in 2022, accounting for 18.6\% of Japan's total exports. Moreover, Japan is the largest foreign holder of Treasury Securities, including Treasury bills, Treasury bonds, and Treasury notes.\footnote{See \url{https://ticdata.treasury.gov/Publish/mfh.txt}} Therefore, the linkages among different financial markets in Japan and the US are of interest for us. 

This study focuses on the period between July 2021 and December 2022. In 2020, the US Federal Reserve implemented a quantitative easing strategy that involved purchasing \$80 billion in Treasury securities and \$40 billion in mortgage-backed securities per month. However, while the US economy was recovering from the Covid-19 pandemic in mid-2021, the Federal Reserve commenced discussions on tapering its asset purchase program. In December 2021, the Federal Reserve initiated the tapering process, reducing monthly asset purchases to support a smooth transition to a post-pandemic monetary policy. In June 2022, the Federal Reserve ended its policy of low interest rates due to concerns over inflation. Meanwhile, the Japanese government maintained its monetary easing policy. In documentation outlining the Bank of Japan’s (BOJ) board meeting in July 2022, policymakers emphasized the need for massive monetary easing to achieve wage increases. As a corollary, the Japanese currency, JPY, weakened throughout 2021 and 2022, reaching a new 32-year low in October 2022. In December 2022, the BOJ announced that it would allow the benchmark 10-year Japanese Government Bond (JGB) yield to move between plus and minus 0.5 percent, instead of the previous tolerable range of plus and minus 0.25 percent. However, the BOJ Governor Mr. Kuroda insisted that it was not an interest rate increase but was intended to “enhance the sustainability of its monetary easing and achieve price stability.”\footnote{See \url{https://www.boj.or.jp/en/about/press/koen_2022/data/ko221226a1.pdf}}  The linkages between the different financial markets in Japan and the US may have changed as a result of divergent monetary policy directions and approaches. Therefore, we apply causal discovery algorithms to data from July 2021 to December 2022, during which the US started the tapering, while Japan continued monetary easing. July 2021 was selected as the start of the time period because the US initiated discussions about tapering in June, which signaled a potential change in monetary policy direction to the markets. We chose December 2022 as the endpoint since it represented the latest available data when we began our research.

\section{Literature Review}
\subsection{Market Linkages}

We begin with an examination of the literature on the linkages among the financial markets of different countries. 

\citet{papana2017financial} applied the conditional Granger causality index and the partial mutual information on mixed embedding to daily data from 1997 to 2015 of 21 stock indices from America, Europe, Asia, and Australia. Both measures suggest that the US stock market has a significant impact on other countries’ stock markets. \citet{saji2022stock} employed the Granger causality test and impulse response function with the daily stock index data of Japan, Singapore, South Korea, India, and China for 1999--2019. This study provides empirical evidence that Asian stock markets have relatively weak price convergence with unidirectional and short-run causality among most markets. \citet{hyvarinen2010estimation} developed VAR-LiNGAM to combine the linear, non-Gaussian, and acyclic model (LiNGAM), with the vector autoregression (VAR) model. Their application of this model to the daily stock market data from December 2001 to July 2006 revealed that US stock affects the following day’s Japanese stock, while the Japanese stock influences the same day’s US stock.

Linkages among bond markets have also become a topic of interest. \citet{jeon2012international} demonstrate that the Japanese bond market is rarely impacted by the bond market of other major industrialized countries, such as the US, the United Kingdom (UK), and Germany, although it does influence markets in other countries. Their study applied the VAR and vector error correction (VEC) models to monthly data on government bond yields of the aforementioned countries from 1980 to 2004. \citet{yang2005international} verified the linkages among government bond markets of five countries, namely the US, Japan, Germany, the UK, and Canada, using monthly data from 1986 to 2000; a recursive cointegration analysis was employed, which revealed that these five markets do not exhibit a long-run relationship. \citet{ito2004financial} compared the interest rate linkages between Japan and the US generated using daily data from two periods: October 1990--May 1993 and May 1993--August 2000. During the former period, Japan and the US both implemented monetary easing, while during the latter period, Japan was still easing and the US was tightening the monetary policy. The cointegration and Granger causality tests show that the uncovered interest rate parity relationship holds true when Japan and the US have similar economic and monetary environments.

Other empirical studies investigate the relationship between the Yen carry trade and related financial markets, such as stock and bond markets. \citet{nishigaki2007relationship} fitted a structural vector autoregression (SVAR) model to monthly data of Japan and the US from 1993 to 2007 with assumptions of the contemporaneous coefficient matrix. This study concludes that the US stock price influences yen carry trades and that yen carry trades affect FX rate. \citet{deaton2020jpy} adopted causal discovery algorithms, LiNGAM, Fast Causal Inference (FCI), and tsFCI to explore the interdependencies among the JPY/Australian Dollar FX rate and the stock and bond markets of the US, Japan, and Australia. The outcomes with 10-minute periodicity data between 2017 and 2018 support the idea that bond markets may not have linkages to the FX rate, while the FX rate can influence some stock and bond markets. 

Previous studies have found evidence of a relationship among the same financial markets and among different financial markets in different countries. Most studies explain the VAR model and Granger causality, which may not correctly handle unobserved variables or instantaneous effects. Two studies have applied causal discovery models focusing on countries and markets different from our research target. Additionally, all previous studies used data before 2019, which did not cover our research period. As the specific conclusions and findings of previous studies vary depending on the country, time period, variables examined, and methodology adopted, we selected the data and methods as suitable for our research topic.

The next section provides an overview of the literature on the methodology based on which we selected the method that was applied in this study.

\subsection{Methods in Causality Research}
\subsubsection{VAR Model and Granger Causality}

VAR, developed by \citet{sims1980macroeconomics} in 1980, is a widely used econometric framework for data description, forecasting, structural inference, and policy analysis. VAR refers to a linear model with $n$ variables and $n$ equations, in which each variable is explained by its own past values, along with the past values of the other variables. This approach enables the structured representation of complex patterns in multiple time series \citep{stock2001vector}. In VAR analysis, it is customary to present findings from various statistical procedures such as Granger causality tests, impulse responses, and forecast error variance decompositions. In particular, Granger causality statistics aim to discern the potential causal relationship between two variables by testing whether a variable’s lagged values are useful for predicting another variable.

However, \citet{stock2001vector} argue that when VAR is used for structural inference and policy analysis, the associated challenges are more complicated because they necessitate distinguishing between correlation and causation. This problem cannot be resolved solely by a statistical approach such as VAR but requires insights from economic theory or institutional knowledge. Moreover, Granger highlighted that the existence of unobserved variables \citep{granger1980testing} or instantaneous effects \citep{granger1988some} may lead to erroneous interpretations of Granger causality.

\subsubsection{Causal Discovery}

The significance of causal relationships lies in the fact that they enable us to predict the behavior of a system when it is subjected to interventions. Such predictions can only be derived by comprehending the fundamental causal relationships underlying the system \citep{eberhardt2017introduction}. Conventionally, identifying causal relationships involves interventions or randomized controlled trials that are often prohibitively expensive, time-consuming, or unfeasible \citep{glymour2019review}. For instance, it is impossible to conduct a controlled experiment to explore the market linkages of financial markets---the topic of this study. Consequently, there has been growing interest in the use of causal discovery, which involves deriving causal information from purely observational data.

Causal discovery aims to determine as much information as feasible about relevant causal relationships from a collection of variables using causal graphs \citep{eberhardt2017introduction}. The process of learning causal models from data always necessitates certain underlying assumptions \citep{heinze2018causal} such as causal sufficiency. The details of the assumptions are provided in Appendix \ref{apd:first}. Given that different causal discovery methods rely on distinct assumptions, the graphical objects differ accordingly (See Appendix \ref{apd:second}).

%\paragraph{2.2.2.3 Causal Discovery Models}\
The \textbf{PC} algorithm, named after its inventors Peter Spirtes and Clark Glymour \citep{spirtes2000causation}, is a constraint-based method with the assumptions of acyclicity, causal faithfulness, and causal sufficiency. Using conditional independence tests, the algorithm aims to infer the structure of the underlying DAG \citep{heinze2018causal}. The three-step process for identifying the completed partially directed acyclic graph (CPDAG) of the DAG involves determining the skeleton, identifying v-structures, and further orienting the edges. The PC algorithm is notable for its ability to extract all possible information regarding the causal structure from conditional independence statements while minimizing the number of tests required \citep{eberhardt2017introduction}. 

The PC algorithm has been subjected to numerous variations and improvements since its inception. One such modification is the \textbf{FCI} algorithm \citep{glymour2019review}. The FCI algorithm weakens the assumptions of the PC algorithm by eliminating the assumption of causal sufficiency and allowing for hidden variables. This weakening of assumptions increases the underdetermination of the underlying causal structure. However, the FCI algorithm conducts additional tests to learn the correct skeleton and has additional orientation rules to ensure completeness. The output of the FCI algorithm can be interpreted as a partially ancestral graph (PAG). Another improved model, the \textbf{PCMCI} algorithm, can handle multivariate and highly auto-dependent time series data, which cannot be managed directly by the PC algorithm \citep{runge2019detecting}. In the first stage of the PCMCI algorithm, the PC algorithm is used to remove most false links while allowing for some false positives. The second stage uses the momentary conditional independence (MCI) test to remove the remaining false links. The PCMCI algorithm assumes causal sufficiency, that is, it does not allow for latent variables. Moreover, the PCMCI algorithm rests on the assumption of stationarity and no contemporaneous causal effects~\citep{runge2019detecting}. The Latent PCMCI (\textbf{LPCMCI}) algorithm  expands the concept of the PCMCI algorithm to apply causal discovery for linear and nonlinear, lagged and contemporaneous causality with latent confounders \citep{gerhardus2020high}. The LPCMCI algorithm improves recall and better controls false positives in the case of autocorrelated time series compared with existing methods, including SVAR-FCI and SVAR-RFCI \citep{gerhardus2020high}. \citet{reiser2022causal} demonstrates that the LPCMCI algorithm outperforms a random algorithm so that it can provide valuable insights when starting with no prior knowledge. 

Although constraint-based methods can generally be leveraged, the fact that the outputs cannot typically be uniquely identified remains a significant disadvantage. With additional assumptions, approaches based on Functional Causal Models \citep{glymour2019review}, which are also known as Structural Equation Models \citep{drton2017structure}, enable the identification of unique DAG. In the \textbf{LiNGAM} algorithm \citep{shimizu2006linear}, the function is linear, and at most one noise term is Gaussian. Independent component analysis \citep{hyvarinen2001independent}, followed by the determination of the correct component ordering through acyclicity, allows us to orient the causal relationship under the assumption of linearity and non-gaussian noise in LiNGAM. The \textbf{VAR-LiNGAM} algorithm \citep{hyvarinen2010estimation} introduced the VAR model into LiNGAM to account for time-series causality. 

\section{Data}

We used daily data for all variables between July 2021 and December 2022, because the Japanese and US markets operate at different times owing to time differences. For the US Dollar (USD)/JPY rate (\textit{USD}), we used data from one of the three Japanese mega-banks, Mizuho Bank. The closing price of the Nikkei 225 index (\textit{Close\_Nikkei}), which represents the Japanese stock market, is obtained from the download center of Nikkei. Nikkei 225 is a stock market index of the Tokyo Stock Exchange that measures the performance of 225 leading publicly traded companies in Japan chosen from various industries. Data on the closing prices of Standard \& Poor’s (S\&P) 500 (\textit{Close\_SP}), a weighted index of 500 listed companies in the US representing the US stock market, and the closing yields of the US 10-year Treasury note (\textit{Close\_US10Y}) were downloaded from the open-source website MarketWatch. Data for the Japanese bond market, including JGB futures closing prices (\textit{Close\_JGBF}) and 10-year JGB closing yields (\textit{Close\_JGB}), were obtained from investing.com, another open-source website. 

We included two different variables for JGB in our study because JGB yield is considered to be closely controlled by the Japanese government, whereas JGB futures prices can illustrate the market more accurately. In fact, the BOJ announced that it would allow the benchmark 10-year JGB yield to move between plus and minus 0.5\% on December 20, 2022, which increased the JGB yield from 0.256\% on December 19 to 0.421\% on that day. 

\section{Method}

To explore the linkages among the FX, stock, and bond markets in Japan and the US between July 2021 and December 2022, we applied causal discovery models to the data introduced in Section 3.
 
First, the original data were visualized using graphs to provide an overview. Second, we adopted the augmented Dickey–Fuller (ADF) test to check whether these time series are stationary, which is an assumption of VAR and VAR-LiNGAM. If the time series are nonstationary, the data will be preprocessed to obtain stationary time-series data. 

The application of the causal discovery algorithms began with VAR-LiNGAM. To apply VAR-LiNGAM to our data, we tested its assumptions. We used the VAR model with all six variables and tested the normality of the residuals of the VAR model by Jarque–Bera test. We additionally verified the linearity of the relationships between the variables using scatterplots. If our verification results do not imply that the assumptions of VAR-LiNGAM are violated, the VAR-LiNGAM can be applied, which in turn will help identify the DAG uniquely. VAR-LiNGAM\footnote{See \url{https://lingam.readthedocs.io/en/latest/tutorial/var.html}}, instead of LiNGAM, is used to consider the time-series impact.

Furthermore, we leverage the LPCMCI model to validate the results when a latent variable exists, as the causal sufficiency assumption is not guaranteed by previous research or related theories. The application of LPCMCI models is realized using Tigramite\footnote{See \url{https://github.com/jakobrunge/tigramite}} package in Python.

\section{Results}
\subsection{Visualization of Original Data}
We first visualized the unprocessed data for the same market, stock, or bond for both Japan and the US in one graph to intuitively perceive the linkages between the two countries. The stock markets of Japan and the US shared a consistent trend for most of the period, as shown in \figureref{fig:stock}, which is consistent with a potential causal linkage between the two. 

\begin{figure}[htbp]
\floatconts
  {fig:stock}
  {\caption{Japanese and US Stock Markets: Nikkei225 and S\&P500}}
  {\includegraphics[width=0.85\linewidth]{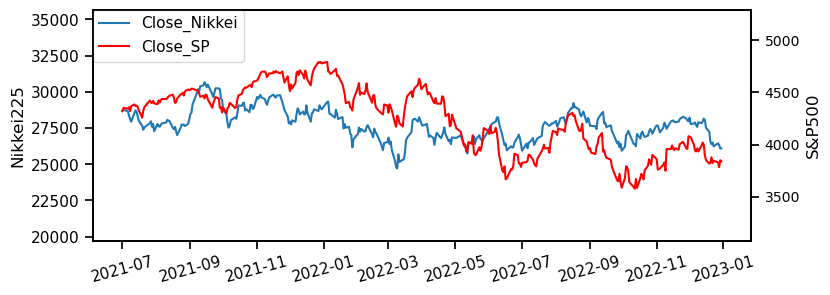}}
\end{figure}

\begin{figure}[htbp]
%\centering
\subfigure[JGB Yield]{
\label{fig:Bond}
\includegraphics[width=7.5cm]{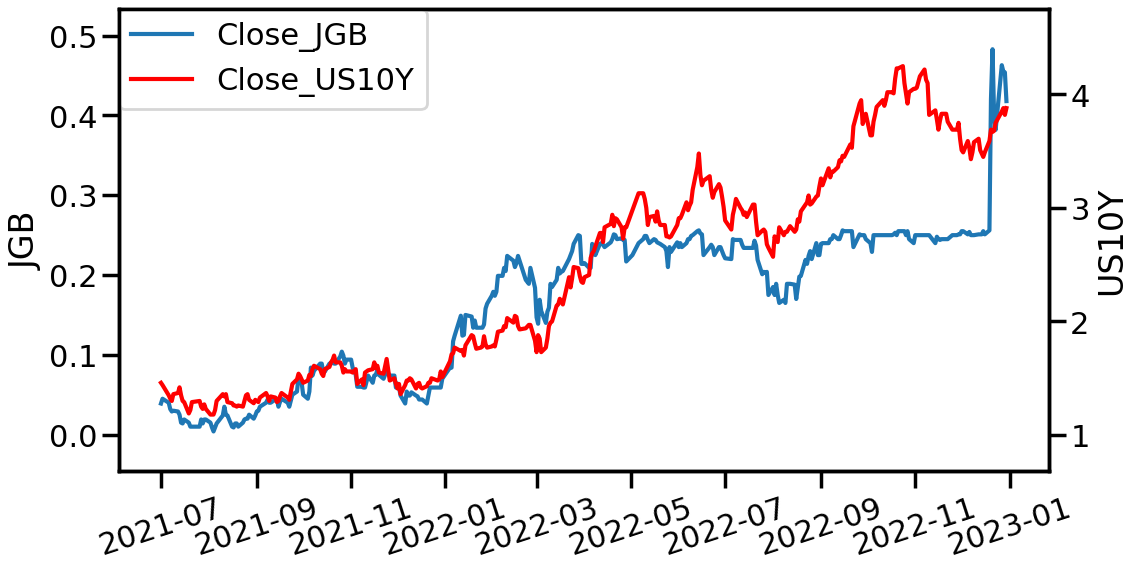}}\subfigure[JGB Futures Price]{
\label{fig:Bond2}
\includegraphics[width=7.7cm]{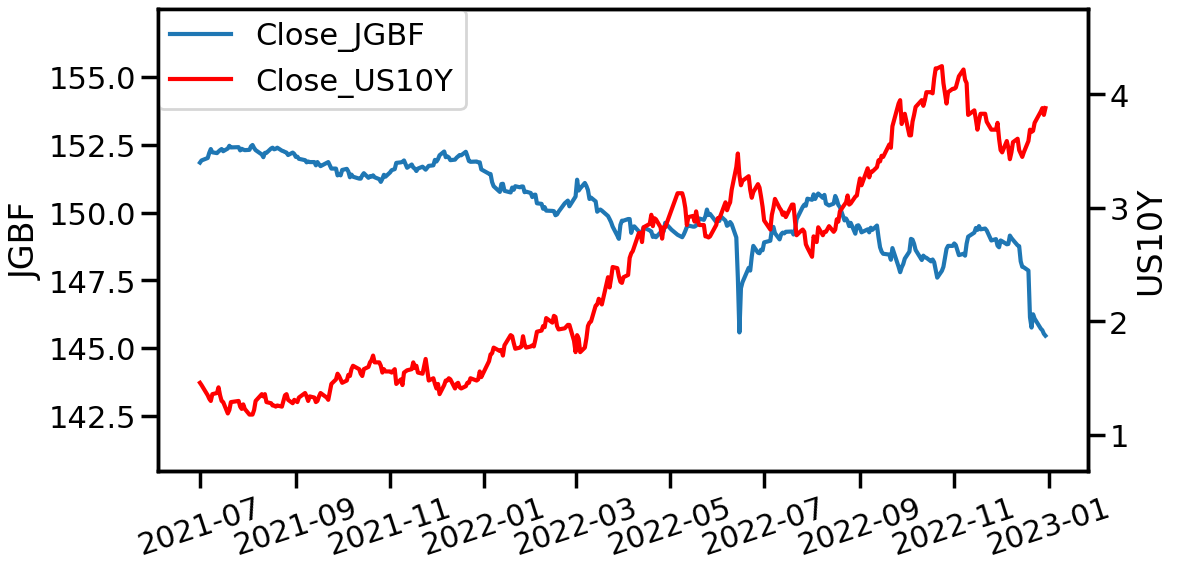}}
\caption{Japanese and US Bond Markets}
\end{figure}

For the bond market, \figureref{fig:Bond} shows a discrepancy between the two countries, especially during the last quarter of 2022, which reflects the Japanese government’s control of the JGB yield. \figureref{fig:Bond2} shows a clear negative correlation between the JGB futures price and the 10-year Treasury note rate. 

\begin{figure}[htbp]
\centering
\subfigure[FX and the US Bond Market]{
\label{fig:FXUSBond}
\includegraphics[width=7.5cm]{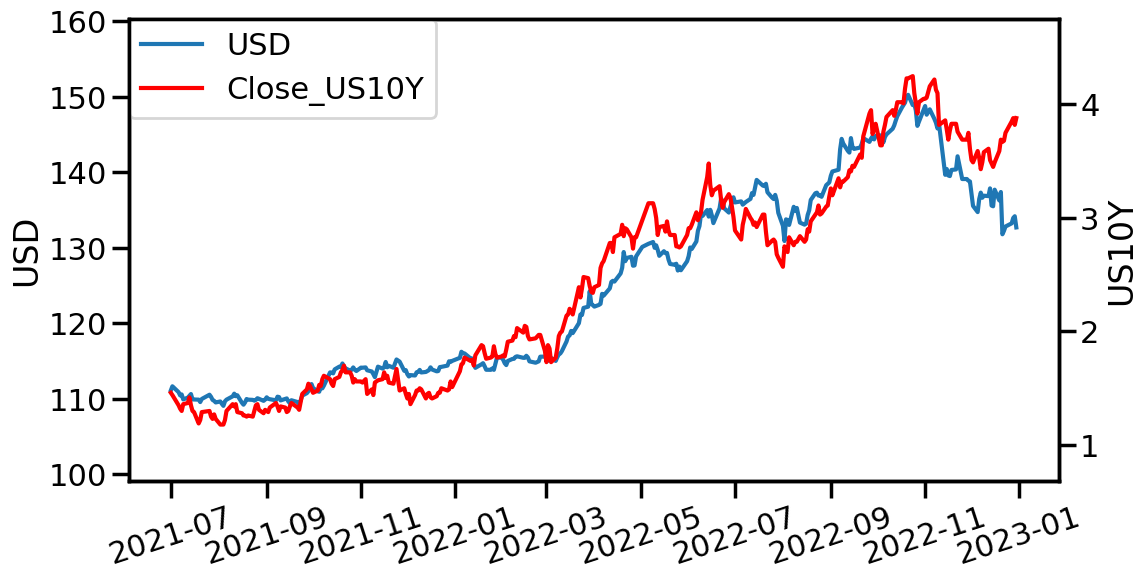}}\subfigure[FX and the US Stock Market]{
\label{fig:FXUSStock}
\includegraphics[width=7.8cm]{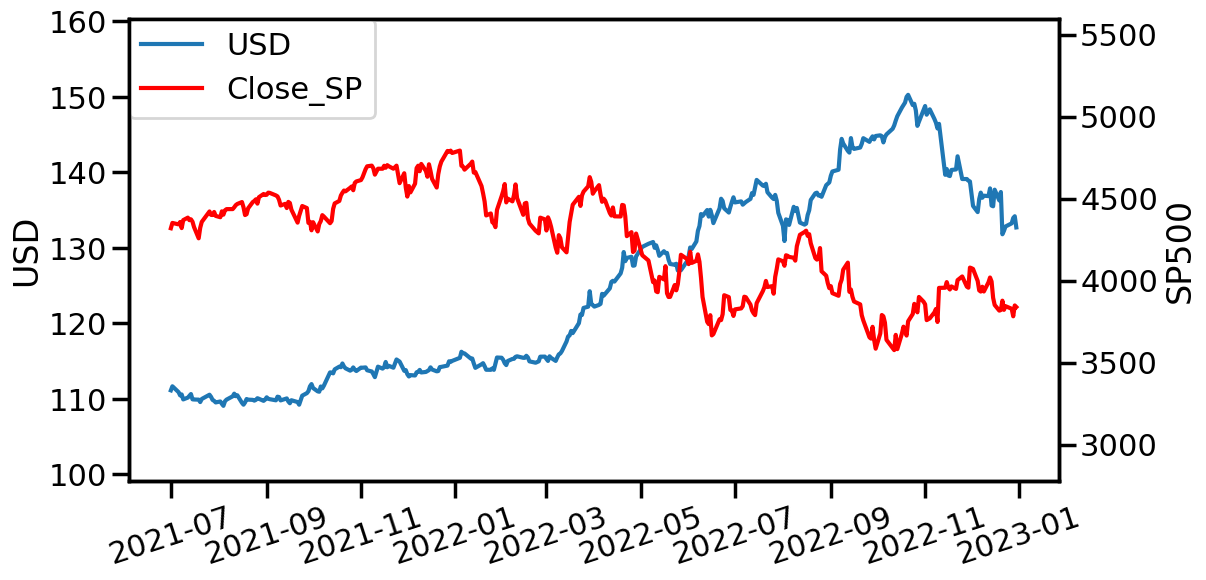}}
\caption{FX and the US Market}
\end{figure}

With regard to the linkages among different markets, the linkage between FX and the US bond market (\figureref{fig:FXUSBond}) is more pronounced than that between FX and the US Stock market (\figureref{fig:FXUSStock}), as the FX rate and the SP500 index are erratically in the same or opposite trend.

\subsection{VAR Model}
As we used VAR for assumption verification and VAR-LiNGAM as one of the methods, the ADF test was adopted to check whether the time series were stationary, which is an assumption of VAR. The results of the ADF test failed to reject the null hypotheses that the time series is nonstationary for all six variables. The $p$-values are larger than 0.05 for all six variables in four different conditions, where “nc” stands for including no constant or trend, “c” stands for including constant only, “ct” stands for including constant and trend, “ctt” stands for including constant and linear and quadratic trends. Therefore, we took their first-order difference and confirmed that the null hypotheses of ADF test are rejected this time, with $p$-values close to zero. The processed data, instead of the original data, are used in subsequent analyses using VAR and causal discovery models. Appendix \ref{apd:third} presents the $p$-values of the ADF test for the original and processed data.

Based on the observations from the visualization, we applied the VAR model to the processed data. Both the Hannan-Quinn information criterion (HQIC) \citep{hannan1979determination} and the Bayesian information criterion (BIC) \citep{schwarz1978estimating} imply that variables at lag 1 should be included in the VAR Order Selection. The results of VAR Model with HQIC are presented in Appendix \ref{apd:fourth}. The results of additional normality (skewness and kurtosis) tests for the residuals of the VAR model rejected the null hypothesis that the residuals follow a normal distribution with a $p$-value close to zero. 

\subsection{VAR-LiNGAM}

In verifying the assumptions for VAR-LiNGAM, we confirmed the non-Gaussian distribution of the disturbance terms in Section 5.2. The scatterplots for each pair of variables (Appendix \ref{apd:fifth}) indicate a linear relationship between some of the variable pairs. In summary, they do not show that our data violates the assumptions of VAR-LiNGAM. We believe it is a good first step toward ensuring the causal sufficiency by including the indices for the three main financial markets---the FX, stock, and bond markets---in the model. Research has demonstrated that the economic environment and monetary policy can influence financial market linkages. However, during the period selected for this study, there was no substantial alteration in either the economic environment or the monetary policies in Japan or the US. Therefore, we applied the VAR-LiNGAM to the data to identify the underlying DAG. The output DAG in \figureref{fig:LiNGAM1} includes the variables at lags 0 and 1, the variables with \textit{(t)}, and the variables with \textit{(t-1)} after order selection. 

\begin{figure}[htbp]
\floatconts
  {fig:LiNGAM1}
  {\caption{Output DAG of VAR-LiNGAM without Domain Knowledge}}
  {\includegraphics[width=1\linewidth]{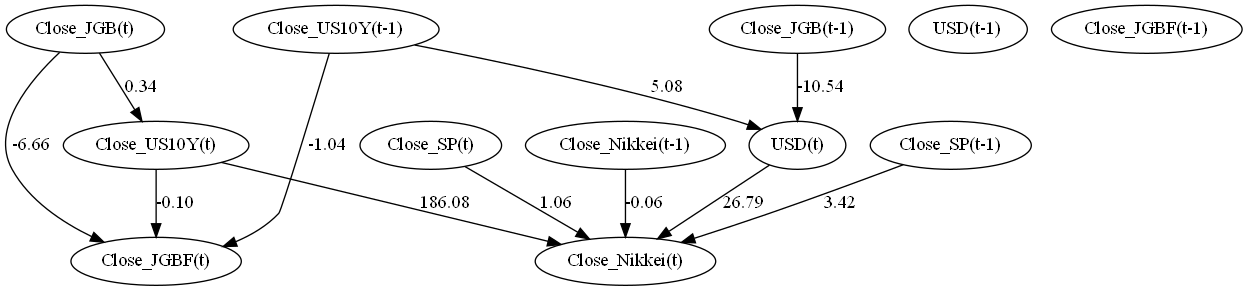}}
\end{figure}

\textit{Close\_US10Y(t)} and \textit{Close\_SP(t)}, the variables at lag 0 representing the US bond and stock markets are illustrated as causally affecting \textit{Close\_Nikkei(t)} and \textit{Close\_JGBF(t)}, the variables representing the Japanese stock and bond markets. Nonetheless, the US markets open after the Japanese markets of the same date close due to the time difference, which leads to the impossibility of directed edges from the US market indices at lag 0 to the Japanese market indices at lag 0. Thus, we exclude these “impossible” directed edges in advance according to the domain knowledge about the markets when adopting the VAR-LiNGAM. The results are shown in \figureref{fig:LiNGAM2}.

\begin{figure}[htbp]
\floatconts
  {fig:LiNGAM2}
  {\caption{Output DAG of VAR-LiNGAM with Domain Knowledge}}
  {\includegraphics[width=1\linewidth]{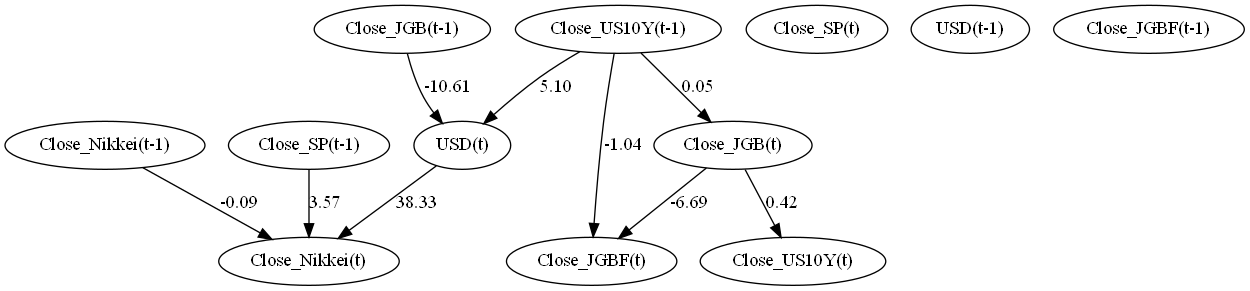}}
\end{figure}

The S\&P 500 index at lag 0 (\textit{Close\_SP(t)}), the JGB futures price at lag 1 (\textit{Close\_JGBF(t-1)}), and the USD/JPY rate at lag 1 (\textit{USD(t-1)}) are not associated with any other variables. The previous business day’s bond market in Japan is an information source for the next day’s USD/JPY rate, because there is a causal flow from \textit{Close\_JGB(t-1)} to \textit{USD(t)}. The previous business day’s US bond market (\textit{Close\_US10Y(t-1)}) is also an information source for the next day’s USD/JPY rate (\textit{USD(t)}), while affecting the next day’s bond markets in both Japan and the US, as illustrated by the arrows from \textit{Close\_US10Y(t-1)} directly to \textit{Close\_JGB(t)} and \textit{Close\_JGBF(t)} and indirectly to  \textit{Close\_US10Y(t)}. The Japanese stock market (\textit{Close\_Nikkei(t)}) functions as an information sink receiving the direct impact from FX market (\textit{USD(t)}), the direct impact from the previous day’s stock markets of both countries (\textit{Close\_Nikkei(t-1)} and \textit{Close\_SP(t-1)}), and the indirect impact from the previous day’s bond markets of both countries (\textit{Close\_JGB(t-1)} and \textit{Close\_US10Y(t-1)}).

The numbers next to the arrows in \figureref{fig:LiNGAM1} and \figureref{fig:LiNGAM2} represent the values in the adjacency matrix calculated from VAR-LiNGAM with the processed data. Since these values depend on the scale of the input data, we calculated the adjacency matrix using standardized data after taking the first-order difference, to understand the strength of causality. (See Appendix \ref{apd:sixth} for the results using standardized data.)

\subsection{LPCMCI}

We leverage the LPCMCI model to validate the results in case a latent variable exists, although the three main financial markets were included and the data period was limited. Similar to our argument in the results of VAR-LiNGAM, the scatterplots indicating the linear relations between variables support the application of LPCMCI model with a conditional independence test for linear association.

In the results of LPCMCI generated by Tigramite package, the graph collapses along the time dimension; thus, each node represents a one-component time series. The edge with the strongest absolute value of the independence test statistic, is displayed between the nodes among all the edges between the two time series. The intensity of the color indicates the strength of causality, with blue indicating a negative impact and red indicating a positive impact. The edge is drawn as a curve with number indicating the lag if there is a lagged causal effect, and as a straight line without number if it is a contemporaneous edge. The numbers on each arrow specify the lags of the causal pathways in the order of their strength.
 
\begin{figure}[htbp]
\centering
\begin{minipage}[t]{0.48\textwidth}
\centering
\includegraphics[width=7cm]{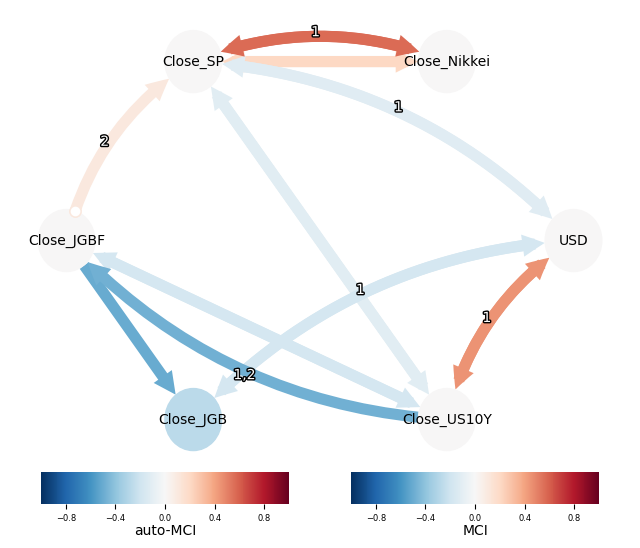}
\caption{LPCMCI with Linear Relation}
\label{fig:LPCMCI1}
\end{minipage}
\begin{minipage}[t]{0.48\textwidth}
\centering
\includegraphics[width=7.3cm]{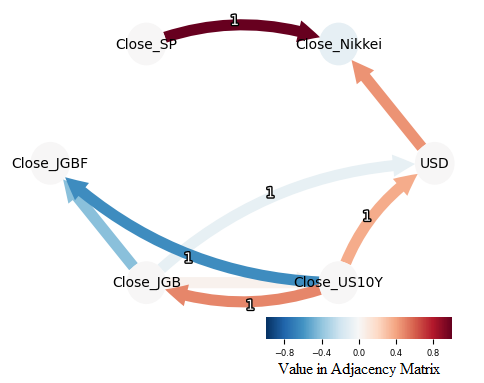}
\caption{VAR-LiNGAM in LPCMCI Form}
\label{fig:LiNGAM3}
\end{minipage}
\end{figure}

As per the results of LPCMCI with linear relation (\figureref{fig:LPCMCI1}), without the causal sufficiency assumption, only the arrow from \textit{Close\_US10Y} to \textit{Close\_JGBF} with number “1, 2” on it remains uniquely directed, which represents the impact of previous one- and two-days’ US bond market on the Japanese bond market. The uncurved arrow from \textit{Close\_JGBF} to \textit{Close\_JGB} at lag 0 indicates that the JGB futures prices causally affects the JGB yield on the same day. Other pathways are either with double arrowheads or open dots and do not provide any prominent insights into the financial markets’ linkages, which might also indicate the existence of latent confounders.

Identical to the procedure that we execute in the adoption of VAR-LiNGAM, we exclude the “impossible” edges, the directed edges from the US market indices at lag 0 to the Japanese market indices at lag 0 known from the time difference of the market operating time when we apply LPCMCI to the processed data. The results of LPCMCI with domain knowledge remains the same as the result without domain knowledge, as shown in \figureref{fig:LPCMCI1}; therefore, only \figureref{fig:LPCMCI1} is included here.

To facilitate a more straightforward comparison of the results obtained from VAR-LiNGAM and LPCMCI, the outcome of VAR-LiNGAM was transformed into the LPCMCI form, as in \figureref{fig:LiNGAM3}. The interpretation of \figureref{fig:LiNGAM3} is the same as LPCMCI, with blue edges representing negative impacts and red edges representing positive impacts. The intensity of the colors represents the value in the adjacency matrix calculated from VAR-LiNGAM with domain knowledge using standardized data after taking the first-order difference, which also indicates the strength of causality. By comparing \figureref{fig:LiNGAM3} with \figureref{fig:LPCMCI1}, we demonstrate that VAR-LiNGAM uniquely identifies the DAG and thus provides informative insight into the causal relationship. For the linkages displayed, both models reveal a causal path from the US bond market (\textit{Close\_US10Y}) to the Japanese bond market (\textit{Close\_JGBF}) at lag 1. However, VAR-LiNGAM identifies the pathway not only to JGB futures price (\textit{Close\_JGBF}) but JGB yield (\textit{Close\_JGB}). Furthermore, VAR-LiNGAM identifies the direction of edges in the bond market (\textit{Close\_US10Y} and \textit{Close\_JGB}), FX market (\textit{USD}), and stock market (\textit{Close\_Nikkei}). Regarding the relationship between JGB futures price (\textit{Close\_JGBF}) and JGB yield (\textit{Close\_JGB}), the two models reach different conclusions regarding the direction of causal relationships, which need further investigation. In addition, the results of LPCMCI contain edges at lag 2, whereas VAR-LiNGAM includes only variables at lag 1 based on the information criterion.

\section{Discussion}

In our study, VAR-LiNGAM was utilized instead of LiNGAM to consider the time-series relationship. LPCMCI is leveraged to handle the possibility of latent confounders because its recall is higher than FCI models in the case of time-series data. The VAR-LiNGAM results is generally consistent with that of LPCMCI, although it uniquely identifies all directions of the edge. No significant causal flows discovered by LPCMCI were missing from the results of VAR-LiNGAM either. By including domain knowledge about the edges into the VAR-LiNGAM in advance, the results were more realistic, particularly for linkages related to bond and stock markets. 

To extend and improve this study further, other variables can be included in the dataset. Although we included three main financial markets and limited the data period, the LPCMCI results disclose the potential of unobserved variables. Variables representing the financial markets of other related countries can be one option, including but not limited to Asian markets, such as the Hang Seng Index in Hong Kong. Another option is to use the sentiment data of market news, which may directly impact several markets simultaneously. For instance, the announcement of governmental intervention in one market may be a common cause of customers’ behavior changes in another.

Moreover, future studies can explore special periods characterized by high market volatility. We observed fluctuations in the trends of our market indices, often accompanied by special events, even when the general macroeconomic environment remains unchanged throughout the selected period. Identifying these high-volatility periods and their contexts will aid in understanding whether they present challenges or opportunities in the application of causal discovery models. Subsequently, this understanding will guide
us in deciding how to handle these specific periods.

Furthermore, the statistical reliability of both VAR-LiNGAM and LPCMCI should be measured to ensure the certainty of the graphs generated. Evaluating reliability will enhance the interpretation of causal graphs from various models, particularly in cases with the inconsistent edge directions, such as between JGB futures price and JGB yield. The bootstrap method can be one of the options to be utilized.

Additionally, researchers can apply the same approach to the data from the beginning of 2020 to the first half of 2021. The results of the two time periods can be compared to confirm whether there is any change in the market linkages when the direction of monetary policy changes.

\section{Conclusion}
\label{sec:theorems}

This study partially elucidates the linkages among different financial markets in Japan and the US during the post Covid-19 pandemic period under divergent monetary policy directions by leveraging causal discovery methods. In particular, integrating domain knowledge in causal discovery is an improved approach compared to previous research in this field, as it helps explore causal relations more accurately and interpret the results more realistically, which is supported by the previous research as well \citep{shen2020challenges}. Although some previous studies deny the existence of linkages related to Japanese financial markets, our results reveal causal pathways among different financial markets and indicate that Japanese markets are an information sink, as shown by other existing studies.

Some results obtained from causal discovery models, especially VAR-LiNGAM, are well-recognized economic phenomena, although they are scarcely verified by empirical data. For instance, the US stock’s previous day closing price affects the following day’s Japanese stock market, and a similar causal flow occurs in bond markets as well. Regarding the linkages among different markets, the VAR-LiNGAM results show that the bond markets of the previous day are the information sources for the following day's FX market, which supports the IRP theory. At the same time, the FX market directly impacts the same day's Japanese stock market. Therefore, bond markets indirectly impact the Japanese stock market the following day through the FX market. This indirect linkage between bond and stock market is not fully explained by economic theories yet. Meanwhile, the LPCMCI results reveal the possibility of latent confounders among the FX, stock, and bond markets in Japan and the US. As noted in Section 6, we consider the financial markets of other related countries, market news that might directly impact several markets, and special events including governmental intervention, as potential unobserved variables.

We believe that understanding the linkages among different markets in different countries will not only help investors make investment decisions with more confidence and less risk, but also help policymakers make more accurate forecasts about their policies’ effect, which was the primary motivation for this study.
 
%\acks{Acknowledgements go here.}
\newpage

\bibliography{pmlr-sample}
\newpage

\appendix
\section{Assumptions of the Causal Discovery Model}\label{apd:first}

\textbf{Causal Sufficiency}. The causal sufficiency assumption pertains to the non-existence of unobserved (or latent) factors \citep{spirtes2000causation}.

\textbf{Acyclicity}. The acyclicity assumption requires that there be no cycles or feedback in the underlying causal model.

\textbf{Causal Faithfulness}. The Causal Markov condition states that every variable in a directed acyclic graph is conditionally independent of its non-descendants given its parents, whereas the causal faithfulness assumption states that all and only the conditional independence relations that hold true in the data are entailed by the Markov condition applied to the graph \citep{glymour2019review}. Essentially, the combination of the causal Markov and faithfulness assumptions suggests that conditional independencies in the probability distribution corresponds to the causal relationships represented by the graph \citep{eberhardt2017introduction}.

\section{Objects of the Causal Discovery Model}\label{apd:second}

Given that different causal discovery methods rely on distinct assumptions, graphical objects differ accordingly \citep{heinze2018causal} \citep{malinsky2016estimating}. 

DAG. The target output of a method with a valid acyclicity assumption is a directed acyclic graph (DAG). A directed edge connects $i$ to $j$ (denoted by $i$ $\rightarrow$ $j$), where $i$ is a direct cause of $j$. A DAG does not allow directed cycles such as $i$ $\to$ $j$ $\to$ 
$i$ based on the acyclicity assumption.

CPDAG. A completed partially directed acyclic graph (CPDAG) condensing the Markov equivalence class of DAGs is obtained in case the DAG is not identifiable with the assumptions and data. $i$ $\rightarrow$ $j$ in a CPDAG indicates that $i$ $\rightarrow$ $j$ exists in every DAG in the Markov equivalence class, whereas $i$ $-$ $j$ indicates that the DAG with $i$ $\rightarrow$ $j$ and $j$ $\rightarrow$ $i$ exist in the Markov equivalence class.

MAG. A maximal ancestral graph (MAG) can be the output of a causal discovery algorithm with the assumption that unobserved variables exist. $i$ $\rightarrow$ $j$ is drawn when $i$ is an ancestor of $j$ and $j$ is not an ancestor of $i$ in the underlying DAG. An edge $i$ $\leftrightarrow$ $j$ is drawn when $i$ is not an ancestor of $j$ and $j$ is not an ancestor of $i$.

PAG. A partial ancestral graph (PAG) can be used if a set of conditional independence relationships can be encoded by multiple MAGs. PAG can have four types of edges: $\rightarrow$, $\circ$$\rightarrow$, $\circ$–$\circ$, and $\leftrightarrow$. The open dot $\circ$ is used for $i$ $\circ$$\rightarrow$ $j$ when $i$ $\rightarrow$ $j$ and $i$ $\leftrightarrow$ $j$ exist in the MAGs.
\newpage

\section{ADF Test $p$-values}\label{apd:third}
\begin{figure}[htbp]
\centering
\subfigure[ADF Test $p$-values for the Original Data]{
\label{fig:ADFOri}
\includegraphics[width=7.5cm]{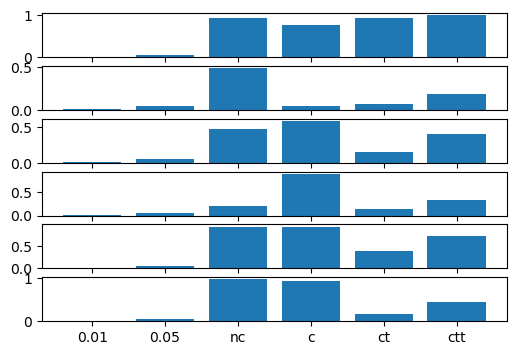}}\subfigure[ADF Test $p$-values for the Processed Data]{
\label{fig:ADFPro}
\includegraphics[width=7.5cm]{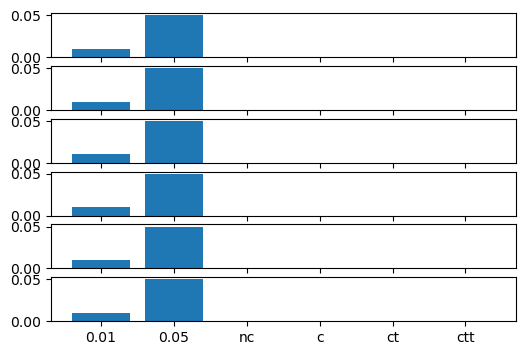}}
\caption*{}
\end{figure}

\section{Results of the VAR Model with HQIC}\label{apd:fourth}
\begin{figure}[htbp]
\centering
\begin{minipage}[t]{0.48\textwidth}
\centering
\includegraphics[width=7cm]{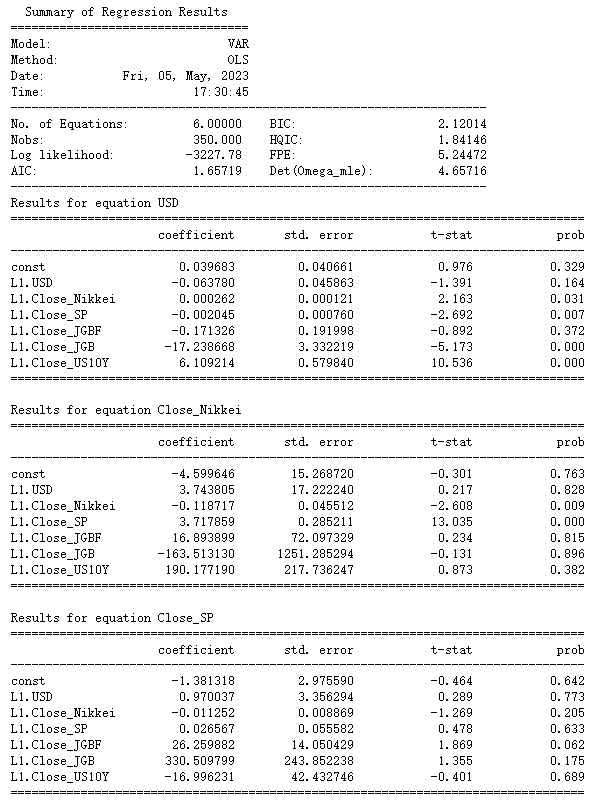}
\caption*{}
\end{minipage}
\begin{minipage}[t]{0.48\textwidth}
\centering
\includegraphics[width=7cm]{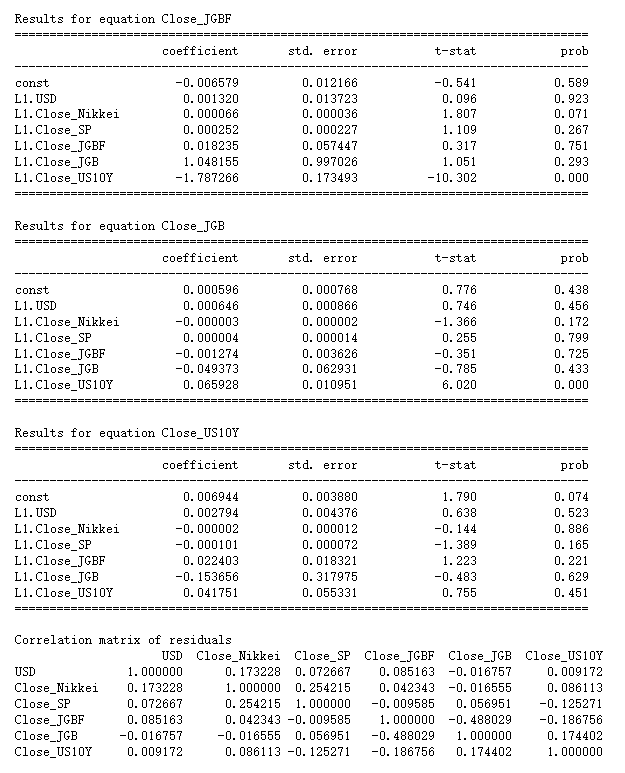}
\caption*{}
\end{minipage}
\end{figure}

\newpage
\section{Scatterplots for Each Pair of the Variables with Lags}\label{apd:fifth}

\begin{figure}[htbp]
\floatconts
  {fig:Corr}
  {\caption*{}}
  {\includegraphics[width=1\linewidth]{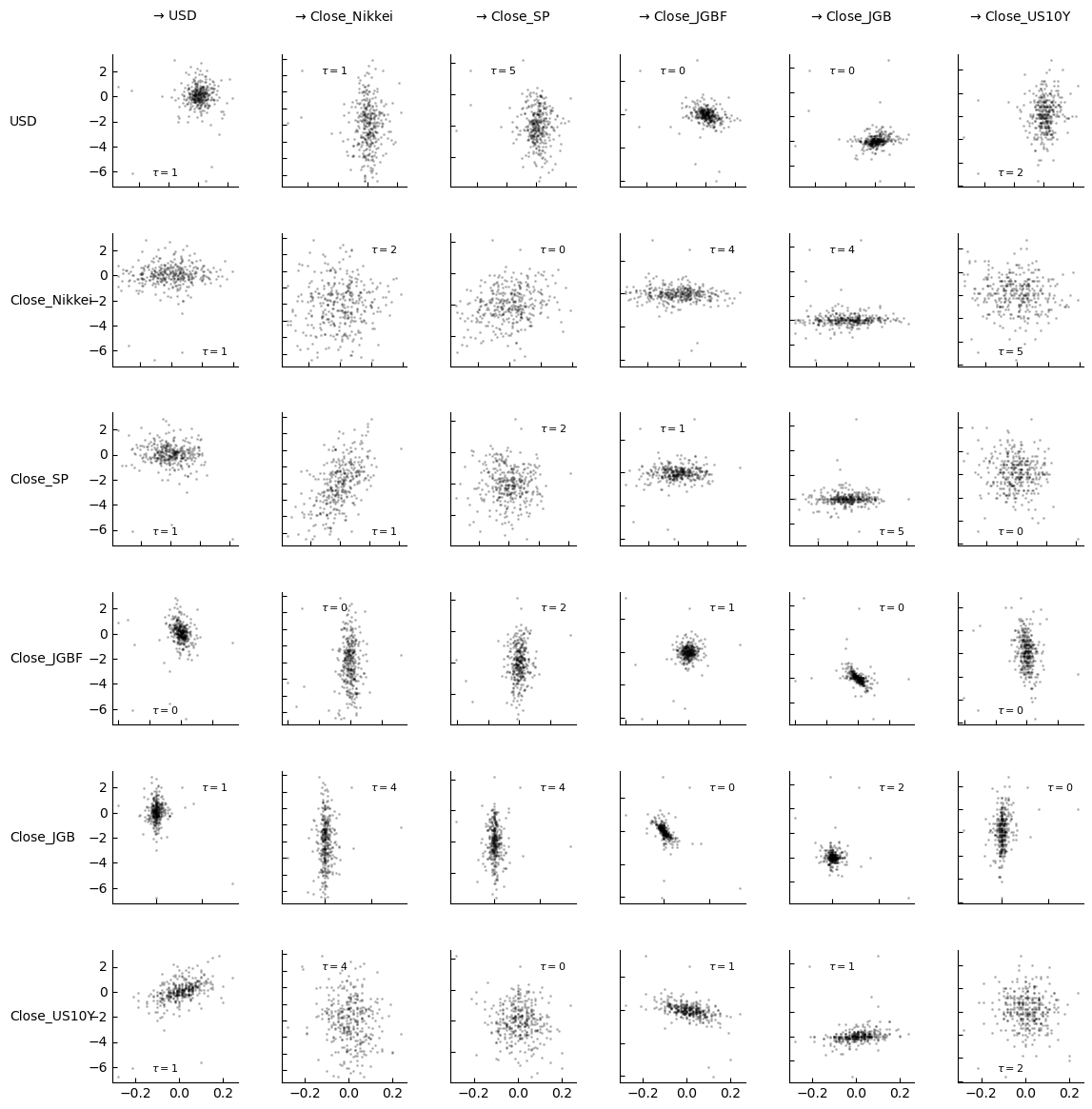}}
\end{figure}

\newpage
\section{Results of the VAR-LiNGAM with Domain Knowledge Using Standardized Data after Taking the First-Order Difference}\label{apd:sixth}

\begin{figure}[htbp]
\floatconts
  {fig:var}
  {\caption*{}}
  {\includegraphics[width=1\linewidth]{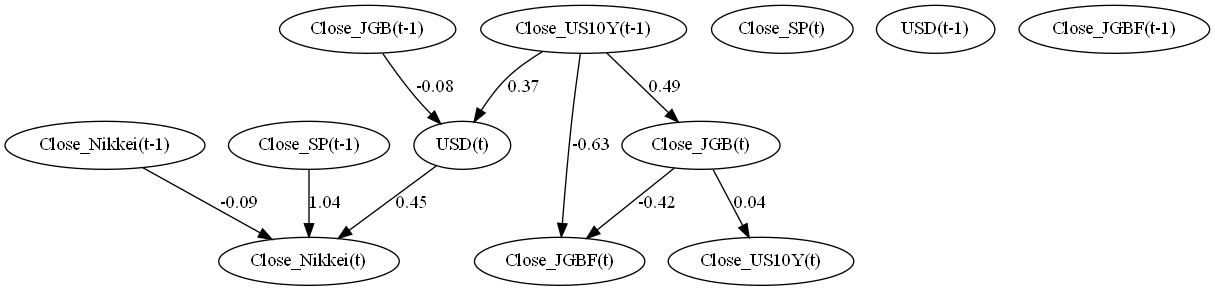}}
\end{figure}

\end{document}